\documentstyle[prd,tighten,aps]{revtex}
\begin{document}

\title{Interplay
of gravitation and linear superposition of different mass eigenstates}

\author{D. V. Ahluwalia$^a$ and C. Burgard$^b$}

\address{$^a$ Mail Stop H--846, Physics (P--25)  Division \\ 
Los Alamos National Laboratory, Los Alamos, NM 87545, USA\\
and\\
Global Power Division, ANSER, S-800\\ 
1215 Jefferson Davis Highway,
Arlington, VA 22202, USA}

\address{$^b$ CERN, PPE,  CH--1211 Geneva 23, Switzerland.}

\maketitle

\begin{abstract}
The interplay of gravitation and the quantum-mechanical principle of linear 
superposition induces a new set of neutrino oscillation
phases.
These ensure that the flavor-oscillation
clocks, inherent in the phenomenon of neutrino oscillations, redshift
precisely as required by Einstein's theory of gravitation.
The physical observability of these phases in the context of the solar neutrino
anomaly, type-II supernovae, and certain atomic systems is briefly discussed.

\end{abstract}

\section{The conceptual framework}

To explain recent atmospheric and solar neutrino data 
the possibility of the violation of the equivalence
principle has been considered \cite{HLP}.
It is therefore an important matter to understand
in detail how quantum mechanics and gravity work together for neutrino 
oscillations. In addition, the problem is of interest in its
own right to understand explicitly, and in detail, how the principle
of equivalence and the principle of linear superposition of quantum mechanics
intermingle. 

The classical effects of gravitation on a single mass eigenstate 
are usually considered in terms of a force $\bbox{F}$, 
whereas the quantum-mechanical ones are determined by the gravitational 
interaction energy.
The interaction energy (denoted by  $U^{{nr}}_{{int}}$) for a 
non-relativistic particle of mass $m$ 
is obtained in the weak-field limit of Einstein's theory of gravitation
to have the same form as in the Newtonian theory \cite{SW}. 
It reads $U^{{nr}}_{{int}}
 = m  \phi $,  with $\phi = -GM/r $ standing for
the gravitational potential of a nonrotating object of mass M, while
 ${\bbox{F}} = - m\,\bbox{\nabla} \phi$. 
The interaction energy for a relativistic
particle $U^{rel}_{int}$  
in the weak--field limit is more generally obtained from the 
force $\bbox{F}$ acting on the mass eigenstate $m$
in the Schwarzschild gravitational environment of a mass $M$  \cite{LBO}:
\begin{equation}
\bbox{F} = -\, {{G M m \gamma}\over {r^3}}\left[
\left(1+\beta^2\right)\bbox{r}-\left(\bbox{r}\cdot\bbox{\beta}\right)
\bbox{\beta}\right]\quad,\label{eq:okun}
\end{equation}
where $\beta={\vert \bbox{v} \vert}/{c} $, 
$\bbox{v}$ is the velocity of the mass eigenstate considered, 
and $\gamma = \left(1-\beta^2\right)^{-1/2}$. 
Assuming the mass eigenstate to be relativistic and setting
$m\gamma=E/c^2$, one is led to the  expression
\begin{equation}
U^{{rel}}_{{int}} 
= \int_\infty^r \bbox{F}\cdot d\bbox{r}^{\,\prime}
=-\, {{GME}\over {r\,c^2}}= - (E/c^2)\,\phi\quad.
\end{equation}
In order to avoid any possible confusion 
we wish to quote Okun explicitly (see \cite{okun}, p.149):
``It is common knowledge that in a locally inertial frame
the gravitational force is zero. That means that [Eq. (\ref{eq:okun})] 
is valid
only for locally [noninertial] frames, such as the usual laboratory frame.''
Now, along an equipotential surface 
the force $\bbox{F}$ is known to vanish and consequently no classical effects 
can be observed in this direction. Indeed,
in going to an appropriately accelerated frame
the constant potential appearing along a segment of an 
equipotential surface can be removed.

To contrast these well-known textbook statements \cite{SW,LBO,okun,JJS}
we wish to focus in the following on some not always fully appreciated 
aspects regarding the phase of a quantum mechanical state. 
A single non-relativistic 
mass eigenstate, considered in an appropriately small 
region of space-time, picks up an unobservable 
global phase factor $\exp\left(-i m c^2 t/\hbar
- i  m \phi t/\hbar \right)$. 
In the case of a linear superposition of $n$ eigenstates 
with different masses ($m_i, i= 1,...,n $), 
their relative phases turn out to be modified through the gravitation
by the mass-dependent factors
$\exp\left(- i m_i \phi t/\hbar \right)$.
We will show below, with appropriate modifications required
for neutrinos, that
these {\it gravitationally induced\/} 
relative phases are observable by a distant observer 
(for local clocks also redshift) as the redshift of the
flavor-oscillation clocks.
To be more specific, during the quantum evolution
of a linear superposition of mass eigenstates the 
corresponding gravitationally induced 
relative phases do not vanish
along an equipotential surface even if 
the gravitational force $\bbox{F}$ does.

For the case considered, 
\begin{equation}
\varphi^G = \Phi \varphi^0\quad,
\end{equation}
where $\Phi=\phi/c^2$ is
the dimensionless gravitational potential, whereas
$\varphi^G$ and $\varphi^0$ stand   
for the (time-)oscillatory
phases with and without gravity, respectively.
 
For neutrino oscillations the gravitationally induced phase 
(denoted by $\varphi^G_{ji}$) 
comes in addition to the corresponding oscillatory phase without 
gravity (denoted by $\varphi^0_{ji}$ ). 
In order to measure the presence of the gravitationally induced phase,
one needs to perform and compare observations on two different equipotential 
surfaces since the experiments on neutrino oscillations
are sensitive only to the sum of both phases.
To extract the full gravitationally induced phase
one of these surfaces should be the surface at spatial infinity. 
Alternatively, it might be instructive
to  compare the above results with those of an experiment performed 
in a freely falling orbiter around the massive object.
Specifically, the
sense in which this comparison is to be performed is identical to that in
which one measures a gravitational redshift of stellar spectra on Earth.
The quantum-mechanically created clocks, via the time oscillation of the
mass eigenstates in the linear superposition, suffer the gravitational
redshift as demanded by Einstein's theory of gravitation
when the gravitationally
induced oscillatory phases are taken into account. 

Theoretically, the prediction of Einstein's theory of gravitation
as regards any clock
(classically or quantum-mechanically driven, 
with a non-relativistic or relativistic mechanism) and the prediction arising
from quantum evolution that incorporates gravity via an interaction
energy term  are in mutual agreement for a nonrotating gravitational
source. The case with a  rotating gravitational source has been 
considered recently  in Ref. \cite{rot}.

In the following, the above-noted 
observations are appropriately modified and applied 
to neutrino weak-flavor eigenstates, which are empirically 
indicated  to be linear superpositions of mass eigenstates.
We again confirm the demands of Einstein's theory of gravitation
in a quantum context.

\section{Redshifting of ``flavor-oscillation'' clocks}

Let us assume that in the ``creation region''
${\cal R}_c$, located at ${\bbox r}_c$,\footnote{The creation region
${\cal R}_c$ is assumed fixed in the global coordinate system.}
 a weak eigenstate with energy $E$, denoted by
$\vert\nu_\ell,\,{\cal R}_c\rangle$, is produced 
with the clock set to $t=t_c$. If each of the three neutrino mass eigenstates
is represented by $\vert \nu_i \rangle$, $i=1,2,3$, then 
one is led to the linear superposition
\begin{equation}
\vert \nu_\ell, \,{\cal R}_c\rangle =
\sum_{i=1,2,3} U_{\ell i}\,\vert\nu_i\rangle\quad,
\label{nu0}
\end{equation} 
where $\ell=e,\mu,\tau$ denotes the  weak-flavor 
 eigenstates as corresponding
to electron, muon, and tau 
neutrinos, respectively.\footnote{
Neutrinos are assumed to be of the Dirac type (for a 
recent analysis of various quantum field theoretic possibilities for the 
description of neutral particles of spin--$1/2$ and higher, and their
relation with space--time symmetries, see Ref. \cite{DVAnp} and references
therein). In addition,
we assume  that
both $\nu_\ell$ and $\nu_m$ are relativistic in the frame of
the experimenter.}
The $\vert \nu_1\rangle$, $\vert \nu_2\rangle$, and 
$\vert \nu_3\rangle$ correspond to the three mass eigenstates
of masses $m_1,\, m_2,\,\,\mbox{and}\,\, m_3$, respectively. Under the 
already-indicated assumptions the 
$3\times 3$
unitary mixing matrix $U_{\ell i}$
may be parametrized by three angles and a CP phase $\delta$
[see \cite{KPbook} Eq. (6.21)].

The evolution of the weak-flavor neutrino eigenstate  
from the creation region ${\cal R}_c$  to the ``detector region'' 
${\cal R}_d$ corresponding to a later time $t=t_d>t_c$
and located at ${\bbox r}_d$
\footnote{Like the creation region, the detection region 
${\cal R}_d$  is also fixed in the global coordinate system.}
is given by the expression
\begin{equation}
\vert{\cal R}_d\rangle =
\exp\left(-\,{i\over {\hbar }} \int_{{t}_c}^{{t}_d}
H \mbox{dt} \,+\, {i\over \hbar}
\int_{{\bbox r}_c}^{{\bbox r}_d}
\bbox P\cdot d\bbox x
\right)
\vert\nu_\ell,\,{\cal R}_c\rangle\quad.\label{gen}
\end{equation}
Here $H$ is the time translation operator
(the Hamiltonian)
 associated with  the system, $\bbox P$ is the operator
for spacial  translations (the momentum operator), and 
$\left[H(t,\bbox x),\,{\bbox P}(t,\bbox x)\right]=0$.
Consider
 that both ${\cal R}_c$ and ${\cal R}_d$ are located in the 
Schwarzschild
gravitational environment \cite{SW}
of a spherically symmetric object of mass $M$.
The direction  of neutrino propagation 
is along $\bbox L = \bbox r_d -\bbox r_c$.
In general, the state $\vert{\cal R}_d\rangle$ is 
not a  definite flavor eigenstate.

As the effects
of astrophysical magnetic fields, and those of the interaction between
the spin of the neutrino and the angular momentum of the astrophysical object
are beyond the scope of the present study
we  neglect the spin--dependent terms for the present. 
It is therefore not necessary to deal with the Dirac equation in the 
curved background. Semiclassical paths and the knowledge that
all energy eigenstates, independent of spin,  carry phases that 
have the general form given by Eq. (\ref{gen}) suffice.
Under these conditions, the exponential on the right-hand side
of Eq. (\ref{gen})
can be evaluated along the line of Stodolsky's work \cite{LS} to
obtain 
\begin{eqnarray}
\exp && \left(-\,{i\over {\hbar }} \int_{{t}_c}^{{t}_d}
H \mbox{dt} \,+\, {i\over \hbar}
\int_{{\bbox r}_c}^{{\bbox r}_d}
\bbox P\cdot d\bbox x
\right)
\vert\nu_i\rangle \nonumber\\
&&\qquad\qquad\qquad\qquad\,=\,
\exp\left[-\,{i\over {\hbar }} 
\int_{{\cal R}_c}^{{\cal R}_d}
\left(
\eta_{\mu\nu}+{1\over 2} h_{\mu\nu}\right)
p_i^{\mu}\,\mbox{dx}^\nu\right]
\vert\nu_i\rangle
\quad .
\label{eq-stodolsky}
\end{eqnarray}
Here $h_{\mu\nu}=g^W_{\mu\nu}-\eta_{\mu\nu}$,
$g^W_{\mu\nu}$ is the 
Schwarzschild space-time metric in the weak-field limit, and $\eta_{\mu\nu}$
is the flat space-time metric. 
In addition, $h_{\mu\nu}=2\,\phi\,\delta_{\mu\nu}$ with the dimensionless
gravitational potential $\Phi= \phi/c^2= -\, G M/c^2 r$.
To avoid notational confusion we remind the reader that $p_i
\equiv \vert \bbox p_i\vert$; the subscript
$i$ identifies the mass eigenstate and does not refer to the $i$th
component of the momentum vector.

We now calculate
the ``neutrino oscillation probability'' from a state
$\vert\nu_\ell,\,{\cal R}_c\rangle$ 
to another state $\vert \nu_{\ell^\prime}, \,{\cal R}_d\rangle$
following closely the standard arguments \cite{KPbook,HL},
appropriately adapted to the present situation.
The oscillation probability
is obtained by calculating the projection
$\langle\nu_{\ell'},\,{\cal R}_d
\vert{\cal R}_d\rangle$ i.e., the amplitude for 
$\vert \nu_\ell,\,{\cal R}_c\rangle \rightarrow
\,\vert\nu_{\ell'},\,{\cal R}_d\rangle$, and then multiplying it by its
complex conjugate. An 
algebraic exercise that  (a) exploits
the unitarity of the neutrino mixing matrix
$U(\theta,\,\beta,\,\psi)$, (b) exploits orthonormality 
of the mass eigenstates, (c) exploits certain trigonometric identities, 
 and  (d) takes care of the fact
 that
now $dx$ and $dt$ are related by
\begin{equation}
dx\simeq\left[1-\left({{2GM}\over {c^2 r}}\right)\right]\,c \,dt
\quad\label{xt}
\end{equation}
yields
\begin{eqnarray}
{\cal P} \Big[\vert\nu_{\ell},{\cal R}_c\rangle\rightarrow 
\vert\nu_{\ell^\prime},{\cal R}_d\rangle\Big] \,= \,
\delta_{\ell\,\ell'}
 &&\,- \,4\,U_{\ell'\,1}\,U^\ast_{\ell\,1}\,U^\ast_{\ell'\,2}\,
U_{\ell\,2}\,\sin^2\left[  
\varphi^0_{21} + \varphi^G_{21}
\right]\nonumber\\
&&\,-\,4\,U_{\ell'\,1}\,U^\ast_{\ell\,1}\,U^\ast_{\ell'\,3}\,
U_{\ell\,3}\,\sin^2\left[
\varphi^0_{31} + \varphi^G_{31}
\right] \nonumber\\
&&\,-\,4\,U_{\ell'\,2}\,U^\ast_{\ell\,2}\,U^\ast_{\ell'\,3}\,
U_{\ell\,3}\,\sin^2\left[
\varphi^0_{32} + \varphi^G_{32}
\right]\quad.
\label{prob}
\end{eqnarray}
The arguments of $\sin^2( )$ in the neutrino
oscillation probability
 now contain two types of phases:
the usual {\em kinematic phase}, denoted here by $\varphi^0_{ji}$
and defined as
\begin{equation}
\varphi^0_{ji} \equiv {{ c^3}\over{ 4\hbar}}
{{ \vert {\bbox r}_d - {\bbox r}_c\vert \Delta 
m^2_{ji}}\over {E}}
=
{{ c^3}\over{ 4\hbar}}
{{L\, \Delta 
m^2_{ji}}\over {E}}
\quad ; \label{phik}
\end{equation}
and the new
 {\em gravitationally induced phase},
denoted here by $\varphi^G_{ji}$ and defined as
\begin{equation}
\varphi^G_{ji} \equiv
{G M c\over{4\hbar}}
\int_{{\bbox r}_c}^{{\bbox r}_d}
{{dl}\over r}
{{\Delta m^2_{ji}}\over E}
\quad, \label{phig}
\end{equation}
to first order in $\phi$ and $\Delta m^2_{ji}$. 
For variations of the gravitational potential that are small compared
to the neutrino energy this integral can be approximated by
\begin{equation}
\varphi^G_{ji} = -\, \langle\Phi
\rangle \,\varphi^0_{ji}\quad,
\label{res}
\end{equation}
where $\langle\Phi\rangle$ is the average dimensionless gravitational
potential over the semiclassical neutrino path
\begin{equation}
\langle\Phi\rangle \equiv
-\,{1\over L}\,\int_{{\bbox r}_c}^{{\bbox r}_d} {dl}\, {{ G M}\over{c^2 r}}
\quad.
\end{equation}

The phenomena of neutrino oscillations provides a 
flavor-oscillation clock.
Substituting
the result (\ref{res}) into Eq. (\ref{prob}), we find 
that  the flavor-oscillation clock redshifts as required by
Einstein's theory of gravitation.

\section{Concluding remarks on the
observability of gravitationally induced phases}

In astrophysical environments 
$\varphi^G_{ji}$ may be a significant fraction of
$\varphi^0_{ji}$. 
However, it is not the ratio
$\varphi^G_{ji}/\varphi^0_{ji}$ alone that
determines the physical relevance of the 
gravitationally induced phases, but also considerations of
{\em astronomical  distances} that are relevant to the problem at hand.

In the
vicinity of a
$1.4$ solar mass neutron star the 
relevant ratio is
\begin{equation}
{{\varphi^G_{ji}} / {\varphi^0_{ji}} }
\approx 0.20\quad.
\end{equation}
Similarly, 
\begin{equation}
\left[\varphi^G_{ji} 
/ 
\varphi^0_{ji}
\right]
_{{Earth}} \sim 10^{-9},\quad
\left[\varphi^G_{ji} 
/ 
\varphi^0_{ji}
\right]
_{{Sun}} \sim 10^{-6},\quad
\left[\varphi^G_{ji}
/ 
\varphi^0_{ji}
\right]
_{{White Dwarf}} \sim 10^{-4}\quad.
\end{equation}
In the gravitational environment of neutron
stars (and hence type II supernovae)
the neutrino oscillation probability is altered at the $20 \%$
level
by the 
gravitationally induced phases and hence cannot be ignored under
most circumstances.
For the propagation of neutrinos in these environments matter effects
may become important. 
These matter effects do not apply equally to all three
flavors of neutrinos and  
modify the Hamiltonian in Eq. (\ref{eq-stodolsky}).
The gravitationally
induced phases are then proportional to the effective kinematic phases,
where the masses are replaced by the effective masses 
in the matter.\footnote{Again, for rapidly changing matter densities this is 
only true differentially and a proper integration has to be performed.}
 The  relative size of the gravitational phases are thus the  same
 as in
the vacuum case. 
For type-II supernovae
the charged current interaction
of $\nu_e$ is important for depositing energy to 
electron-flavor rich matter. This energy deposit is  
dramatically modified by neutrino oscillations including their 
gravitational modifications if the $\Delta m^2_{ji}$ are such as to yield
oscillations lengths that match relevant length scales for supernovae.

The effect of the gravitationally induced phases on the oscillation
probabilities also depends on the number of oscillation between ${\cal R}_c$ and
${\cal R}_d$. For the
gravitational field of the Sun, for example, a gravitational phase of
$\pi$ is accumulated after $5\times 10^{5}$ oscillations. Whether or not
the oscillations are washed at this point depends on the energy spread and
the size of the source (and the detector).

In the 
latest neutron interferometry experiments a discrepancy between theory and
experiments continues to exist at the several standard deviation level 
\cite{COWd}.
Therefore, the  question naturally arises if there exist other
physical systems where the interplay of quantum mechanics and gravitation
may be studied for a better understanding of the experimental systematic errors
involved.
Towards this end we note that
it  remains possible to study atomic systems that are in a linear
superposition of two or more energy states in terrestrial experiments.
In such systems any violation of the equivalence principle at a level
of 1 part in $10^9$ or study of 
the existence of the gravivector and
graviscalar fields  that arise in supergravity \cite{sb}
becomes experimentally accessible.\footnote{After this manuscript 
was completed we 
learned of a precision atomic interferometry experiment \cite{atomic}  
where a variation
in Earth's 
gravity due to tides at Stanford was measured via gravitationally induced phases.
Such experiments, when done with controlled local gravity sources, rather than tides,
open the study of deviations \cite{dev} in the predictions based upon the 
principle of equivalence and the principle of linear superposition.}

In summary, the basic result of this paper is as follows.
The phenomenon of neutrino oscillations 
provides a ``flavor-oscillation clock.''
The flavor-oscillation clock
redshifts as required by Einstein's theory of gravitation.
Apart from the 
fact that these results are important for type-II supernova evolution,
one may
also study the interplay of the 
principle of equivalence  and the quantum mechanical 
linear superposition in atomic systems with existing technology. The
integration of the gravitationally induced phase over the
Earth-Sun distance may be relevant for the solar neutrino anomaly.

{\bf Acknowledgments}

We are indebted to Steven Weinberg for a comment on an earlier version of
this work that led to a deeper
understanding of the interplay of gravitation and quantum mechanics
and Sam Werner for providing us with a copy of Ref. \cite{COWd}
before its publication. 
We thank Ephraim Fischbach for bringing to our attention
Ref. \cite{ef} where gravitational effects 
on $K^0$-$\overline{K}^0$ systems have been considered.
We further acknowledge useful conversations with Sam Werner and 
Ephraim Fischbach. Continuing discussions on space--time symmetries
with V. Raatriswapan are gratefully acknowledged.
This work was done, in part, under the auspices of the 
U. S. Department of Energy.

\end{document}